\documentclass[structabstract]{aa}  

\usepackage{graphicx}
\usepackage[varg]{txfonts}
\usepackage{amsmath}
\usepackage{longtable}
\usepackage{longtable,lscape}
\usepackage{lscape}
\usepackage{epsfig}

\newcommand\be{\begin{equation}}
\newcommand\en{\end{equation}}

\begin{document} 

\title{A Herschel view of IC\,1396\,A:\\ Unveiling the different sequences of star formation\thanks{Based on observations obtained with the
{\it Herschel} Space Telescope within Open Time proposal "Disk dispersal in Cep OB2", OT1\_asicilia\_1. {\it Herschel} is an ESA space
observatory with science instruments provided by European-led PI consortia and with important participation from NASA.}}
 
%\subtitle{Disk evolution, accretion, and environment}

\author{Aurora Sicilia-Aguilar\inst{1,2}, Veronica Roccatagliata\inst{3}, Konstantin Getman\inst{4}, Thomas Henning\inst{5},\\Bruno Mer\'{\i}n\inst{6}, Carlos Eiroa\inst{1}, Pablo Rivi\`{e}re-Marichalar\inst{7}, Thayne Currie\inst{8}}

\institute{	\inst{1}Departamento de F\'{\i}sica Te\'{o}rica, Facultad de Ciencias, Universidad Aut\'{o}noma de Madrid, 28049 Cantoblanco, Madrid, Spain \\
	\email{aurora.sicilia@uam.es}\\
	\inst{2}SUPA, School of Physics and Astronomy, University of St Andrews, North Haugh, St Andrews KY16 9SS, Scotland, UK\\
	\inst{3}Universit\"ats-Sternwarte M\"unchen, Ludwig-Maximilians-Universit\"at,Scheinerstr.~1, 81679 M\"unchen, Germany \\
	\inst{4}Department of Astronomy \& Astrophysics, 525 Davey Laboratory, Pennsylvania State University, University Park, PA 16802, USA\\
	\inst{5}Max-Planck-Institut f\"{u}r Astronomie, K\"{o}nigstuhl 17, 69117 Heidelberg, Germany\\
	\inst{6}Herschel Science Centre, ESAC-ESA. PO Box 78. E-28691 Villanueva de la Ca\~{n}ada, Madrid, Spain \\ 
	\inst{7}Kapteyn Astronomical Institute, P.O. Box 800, 9700 AV Groningen, The Netherlands\\
	\inst{8}Department of Astronomy \& Astrophysics, University of Toronto, Canada\\}
	
   \date{Submitted September 5, 2013, accepted December 11, 2013}

\abstract
{The IC\,1396\,A globule, located to the west of the young cluster Tr\,37, is known to host many very
young stars and protostars, and is also assumed to be a site of triggered star formation.}
%aims
{Our aim is to test the triggering mechanisms and sequences leading to star formation in Tr\,37
and similar regions.}
%method
{We mapped IC\,1396\,A with Herschel/PACS at 70 and 160\,$\mu$m. The maps reveal the structure of the
most embedded parts of the star-forming site with great detail.}
%results
{The Herschel/PACS maps trace the very embedded protostellar objects and the structure of the
cloud. PACS data reveal a previously unknown Class 0
object, labeled IC1396A-PACS-1, located behind the ionization front. IC1396A-PACS-1 is 
not detectable with Spitzer, but shows marginal X-ray emission.
The data also allowed to study three of the Class I intermediate-mass objects within the cloud.
We derived approximate cloud temperatures to study the effect and potential interactions
between the protostars and the cloud. The Class 0 object is associated with the
densest and colder part of IC\,1396\,A.  Heating in the cloud
is dominated by the winds and radiation of the O6.5 star HD\,206267 and, to a lesser extent, by the effects 
of the Herbig Ae star V 390 Cep. The surroundings of the Class I and Class II objects embedded in the cloud
also appear warmer than the sourceless areas, although most of the low-mass 
objects cannot be individually extracted due to distance and beam dilution.}
%conclusions
{The observations suggest that at least two episodes of star formation have occurred in IC\,1396\,A. One
would have originated the known, $\sim$1 Myr-old Class I and II objects in the cloud, and a 
new wave of star formation would have produced the Class 0 source
at the tip of the brigth-rimmed cloud. From its location and properties, IC1396A-PACS-1 is consistent with having
been triggered via radiative driven implosion (RDI) induced by HD\,206267.
The mechanisms behind the formation of the more evolved population of Class I/II/III objects in the cloud are 
uncertain. Heating of most of the remaining cloud by Class I/Class II
objects and by HD\,206267 itself may preclude further star formation in the region.}


\authorrunning{Sicilia-Aguilar et al.}

\titlerunning{Herschel observations of IC1396~A}

\maketitle

%________________________________________________________________

\section{Introduction \label{intro}}

The IC\,1396\,A globule is part of one of the most remarkable H~II regions in the
northern hemisphere. Located at the western edge of the Tr\,37 cluster (Marschall \& van Altena 1987;
Platais et al. 1998) at 870 pc distance (Contreras et al. 2002),
it appears as a blown-away structure shaped by the stellar winds of the massive
stars in Tr\,37. The Trapezium-like system HD\,206267, dominated
by a O6.5 star (Abt 1986; Peter et al. 2012), is thought to be the main source of ionization of 
IC\,1396\,A, which is located at about 4.5\,pc projected distance to the west of HD\,206267.

\begin{figure*}
\centering
\epsfig{file=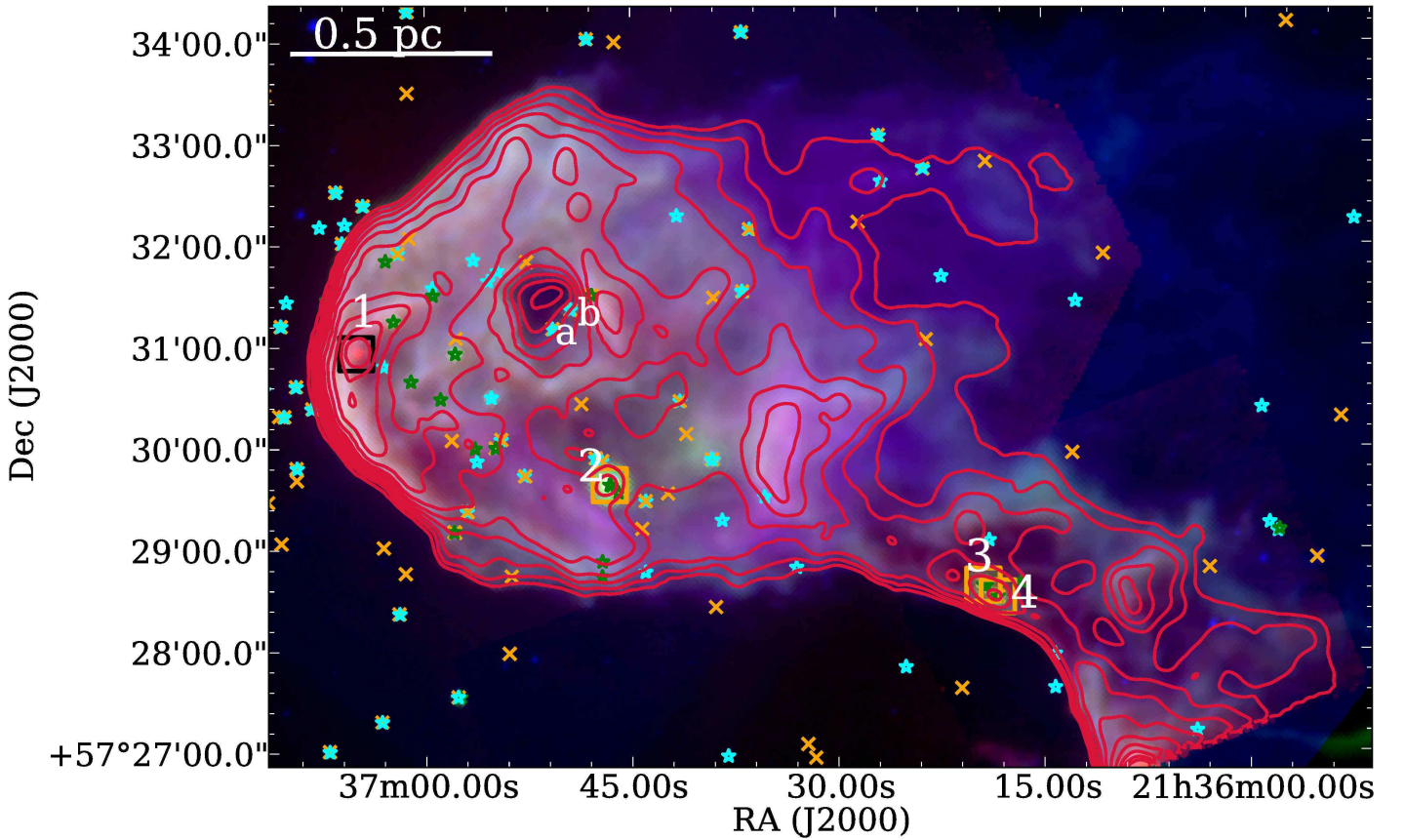,width=1.0\linewidth,clip=}
\caption{A 3 color (8, 24, 70\,$\mu$m image of the IC\,1396\,A globule. The cyan stars denote Class II and Class III objects
from Sicilia-Aguilar et al. (2005, 2006a,b), Morales-Calder\'{o}n et al. (2009), and Getman et al.(2012). The green
stars are Class I objects from Sicilia-Aguilar et al.(2006a) and Reach et al. (2004). The yellow X symbols correspond
to X-ray detections consistent with YSO from Getman et al. (2012). The objects discussed in this work are marked as
large squares (black for Class 0 objects, orange for Class I objects) and labeled with numbers (1=IC1396A-PACS-1;
2=21364660+5729384; 3=21361942+5728385; 4=21361836+5728316). We also labeled the objects in the eye-shaped
hole of the nebula (marked as cyan stars; a=V~390~Cep; b=14-141, see text). 
The contours correspond to the 160\,$\mu$m PACS image, 
10 levels in log scale between 0.7 and 10 Jy/beam.   \label{3color-fig}}
\end{figure*}

Tr\,37 is part of the Cep\,OB2 region (Simonson \& van Someren Greve 1976),
which is considered as an example of triggered or sequential star formation (Elmegreen 1998). 
Loren et al. (1975) found strong molecular line emission associated with IC\,1396\,A, suggestive 
of material with different column densities and temperatures,
and postulated the existence of embedded sources that would contribute to the heating of the cloud. 
IC\,1396\,A is a bright-rimmed cloud (BRC; Sugitani et al. 1991; Sugitani \& Ogura 1994) 
that is believed to be suffering dispersal
and collapse due to triggered star formation induced by the massive and 
intermediate-mass stars in Tr\,37. Sugitani et al. (1991) proposed the presence of 
a triggered population of young stellar objects (YSOs) with IR excesses 
identified with IRAS. The star-formation sequence would have started with
the 12 Myr old cluster NGC~7160, giving later on rise to the younger (4 Myr) Tr\,37 
cluster and associated structures (Patel et al. 1995, 1998). Patel et al. (1995) CO 
observations also revealed a column density enhancement shaped by the ionization 
front of the BRC. The CO and NH$_3$ studies of
Morgan et al. (2009, 2010) are also consistent with ongoing, probably triggered,
star formation as the origin of the Class I
object at the edge of the southern part of IC\,1396\,A (21360798+572637/$\gamma$;
Sicilia-Aguilar et al. 2006; Reach et al. 2004). Nevertheless, conclusive evidence of
triggered star formation based on gas kinematics is not always easy to obtain 
(e.g. Chen \& Huang 2010; Mookerjea et al. 2012), since dynamical evolution may contribute 
to erase the signatures of triggered star formation with time.

The Spitzer Space Telescope revealed in great detail a rich, embedded population of 
low-mass stars and protostars within IC\,1396\,A. Spitzer unveiled about 60 very embedded 
sources, mostly Class I/II, plus a similar number of low-mass and solar-type Class II/III 
objects located within $<$4' (1 pc at 870 pc distance) of the globule
(Reach et al. 2004; Sicilia-Aguilar et al. 2006a, 2013b; Morales-Calder\'{o}n et al. 2009). 
Chandra X-ray observations revealed 250 objects in and near the BRC, doubling
the previously known population down to masses $\sim$0.1~M$_\odot$ (Getman et al. 2012). 

The stellar population associated with IC\,1396\,A is clearly differentiated
in isochronal age. Not only the objects in-cloud appear to be younger, 
but there is also evidence for a spatial age gradient with younger stars 
($\le$1 Myr, according to the Siess et al. 2000 isochrones) appearing in an arc-shaped region
around IC\,1396\,A (Sicilia-Aguilar et al. 2005). X-ray detected sources also show isochronal
age differences, with objects near the globule being as young as 1-2 Myr, while the main cluster
population has typical ages of 4 Myr (Getman et al. 2012). There is also a spatial 
evolutionary-state gradient. While optical and Spitzer studies of the main Tr\,37 cluster revealed 
only Class II and Class III sources, with a disk fraction
about 48$\pm$5\% and disk IR excesses lower than Taurus (Sicilia-Aguilar et al. 2006a, 2013b), 
IC\,1396\,A contains Class II objects with Taurus-like disks and several embedded Class I 
sources (Reach et al. 2004; Sicilia-Aguilar et al. 2006a; Morales-Calder\'{o}n et al. 2009).

The age gradient observed from the ionizing star HD\,206267 towards the cloud 
suggested star formation due to radiative driven implosion (RDI) that
could have lasted for several Myr. In the RDI scenario, the ionization front from OB stars 
ablates the surface of surrounding cloudlets, producing cometary structures and driving 
inward a compression shock that induces star formation. Discussed since the 1980s, RDI has 
now well-developed hydrodynamical calculations and predictions (Kessel-Deynet \&
Burkert 2003; Miao et al. 2008).  RDI could be responsible for a substantial part ($\sim$14-25\%)
of the star formation in the region (Getman et al. 2012). 
Narrow-line imaging also reveals signs of ionization at the cloud edge 
(Sicilia-Aguilar et al. 2004, 2013b; Barentsen et al. 2011), showing that not only
HD\,206267 has an effect on the cloud, but the low-mass T Tauri stars in the 
region also appear to interact with their surroundings, contributing to the 
removal and dispersal of the cloud material (Sicilia-Aguilar et al. 2013b). 
The most remarkable example may be the cleared, eye-like hole in the
center of the globule, created by the intermediate-mass
star V 390 Cep and the classical T Tauri star 14-141 (see Figure \ref{3color-fig}).

Here we present the Herschel/PACS observations of IC1396~A.
PACS scan maps at 70 and 160\,$\mu$m were obtained as part of the Open Time Program 
"Disk dispersal in Cep OB2" (PI A. Sicilia-Aguilar). 
In the present work we concentrate on the star formation and cloud structure 
of IC\,1396\,A. More detailed discussion on the stars with disks in the Cep OB2 region,
including IC\,1396\,A and surroundings, will follow in a coming publication.
The observations and data reduction are described in Section \ref{data}.
Section \ref{analysis} presents the analysis of the point sources and cloud structure. 
The discussion of the implications for the star formation history in the region 
are presented in Section \ref{discussion}, and Section \ref{summary} summarizes our results.

\section{Observations and data reduction\label{data}}

The IC\,1396\,A globule was observed with the ESA Herschel Space Observatory (Pilbratt et al. 2010) using 
the Photodetector Array Camera and Spectrometer (PACS; Poglitsch et al. 2010).
The observations comprised a large scan map (AORs 1342259791 and 1342259792)
and a small mini-map field (AORs 1342261853 and 1342261854). They were executed
by Herschel on 2013-01-16 and 2013-01-23, respectively. The two AORs of the large map 
correspond to scan and cross-scan (at 45 and 135 degrees with respect to the array) 
of a 13.5'$\times$13.5' field centered around 21h38m18.770s +57d31m36.80s at 70 and 160\,$\mu$m. 
The mini-map field was observed at the same wavelengths, centered on 21h36m25.080s +57d27m50.30s,
and consisted of a scan and cross-scan at 70 and 110 degrees with respect to the array.
The field was selected to contain the main part of the IC\,1396\,A globule, together with the 
younger disk population in Tr\,37 (the disk population of Tr\,37 will be discussed in a 
separate paper). The large scan map was executed in 1.8h, and the small mini-map
required 20 minutes of Herschel time, both using the medium (20"/sec) scan speed. 
The final sensitivity 
of the image depends strongly on the presence of extended structures, resulting
in a lower contrast in the areas dominated by the IC\,1396\,A globule emission, than in the 
clean areas where the older, Class II/III Tr\,37 population is located.

The data were reduced using 
{\sc HIPE}\footnote{{\sc HIPE} is a joint development by the Herschel Science Ground Segment 
Consortium, consisting of ESA, the NASA Herschel Science Center, and the HIFI, 
PACS and SPIRE consortia.} environment, version 11.0 (Ott et al. 2010) 
and the Unimap\footnote{See http://w3.uniroma1.it/unimap/ for further details} 
software (Piazzo et al. 2012). The PACS data calibration is that of 
2013 February 14. The basic corrections (up to level 1) were done with {\sc HIPE}. 
The {\sc HIPE} reduction was based on the standard Bright Source {\sc HIPE} templates
up to level 1.  It included identifying the science frames, flagging the bad and saturated 
pixels, converting from ADUs to Volts, converting chopper angles to sky angles, computing the
frame coordinates, and calibrating the data. The main complication of the region is the failure 
of the standard {\sc HIPE} high-pass filtering (HPF) in a field where the
extended emission is the main science target, so the final mapping was done with Unimap. 
First, the Unihipe interface was used with {\sc HIPE} to produce the Unimap input files, which
for PACS data are essentially the level 1 products.
Unimap is based on the Generalized Least Square (GLS)
mapping technique, but has additional features to remove offsets, cosmic rays, calibration and baseline drifts,
and glitches (Piazzo et al. 2012). To take care of the distortion and noise resulting from the GLS
processing, Unimap includes two post-GLS routines for removal of distortion and minimization of the map noise.
After creating a naive map (containing signal and correlated noise) based on the level 1 {\sc HIPE} results, 
Unimap searches for and flags signal jumps.
Later on, it detects, flags, and corrects glitches applying a HPF of 25". Drifts
are corrected by a Subspace Least Square (SLS) technique. Noise corrections  are applied by estimating the noise
power and spectrum. Then the GLS map routine is applied, and if the distortion is stronger than the correlated noise,
the map is further corrected, estimating the distortion by a median filtering and naive projection. This 
correction tend to increase the noise of the image, which is finally reduced applying a weighted noise correction.
The final map is projected with pixel sizes 2" for the 70\,$\mu$m image, and 3" for the 160\,$\mu$m image.

\begin{figure*}
\centering
\epsfig{file=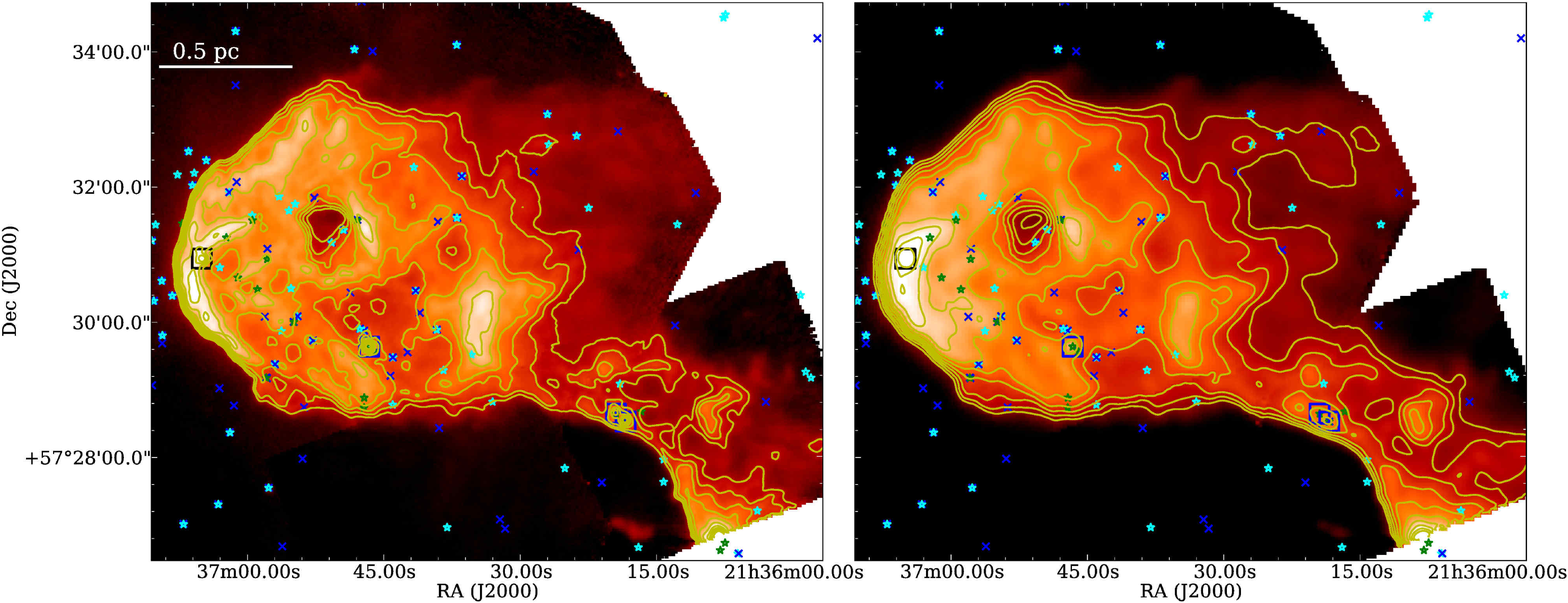,width=1.0\linewidth,clip=}
\caption{IC\,1396\,A as seen at 70 (left) and 160\,$\mu$m (right). The cyan stars denote Class II and Class III objects
from Sicilia-Aguilar et al. (2005, 2006a,b), Morales-Calder\'{o}n et al. (2009), and Getman et al.(2012). The green
stars are Class I objects from Sicilia-Aguilar et al.(2006a) and Reach et al. (2004). The blue X symbols correspond
to X-ray detections consistent with YSO from Getman et al. (2012). The objects discussed in this work are marked as
large squares (black for Class 0 objects, blue for Class I objects). The contours mark 10 levels in log scale
from 0.1-2.0 Jy/beam (70\,$\mu$m) and 0.7-10 Jy/beam (160\,$\mu$m).  \label{70um160um-fig}}
\end{figure*}

At present, Unimap does not allow to combine large scan maps and mini-maps. For photometry purposes, we used
the individual maps, which involve less post-processing. To study the whole cloud structure, a combined, 
mosaicked map of the four AORs was constructed using the new mosaicking tool
within {\sc HIPE} version 11.0. We first used {\sc HIPE} on the mini-map to rotate it, trim its edges
(to avoid an increase in the noise at the map edges affected by low coverage), and derotate it. We finally
combined the rotated mini-map and the larger scan map with the mosaicking tool. 
Due to the different orientations of the images, the final mosaic is
resampled by {\sc HIPE} to 1/3 of the original pixel size. The final mosaicked maps around the IC\,1396\,A 
region are displayed in Figures \ref{3color-fig} (for a 3-color map including Spitzer data) and \ref{70um160um-fig} 
(for a detailed view of the Herschel/PACS data at both wavelengths).

\section{Analysis \label{analysis}}

\subsection{Source photometry and luminosities \label{photometry}}

\begin{figure*}
\centering
\epsfig{file=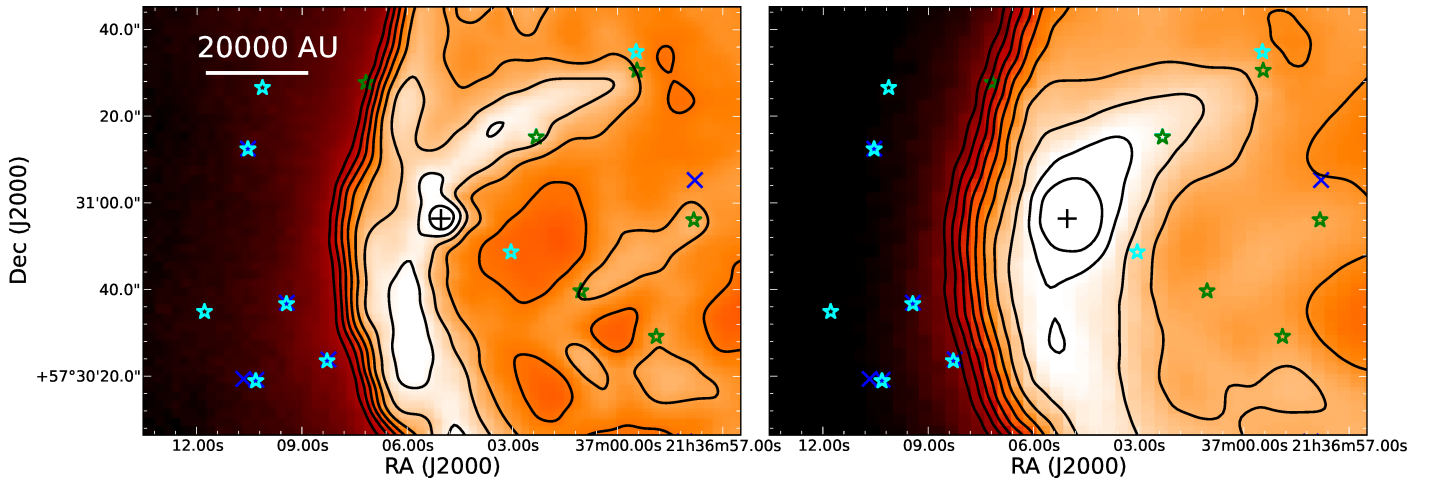,width=1.0\linewidth,clip=}
\caption{A zoom in the surroundings of the newly discovered source IC1396A-PACS-1 at 70\,$\mu$m (left) and 160\,$\mu$m (right).
The black '+' marks the center of the new source, according to the 70\,$\mu$m map. 
The cyan stars denote Class II and Class III objects
from Sicilia-Aguilar et al. (2005, 2006a,b), Morales-Calder\'{o}n et al. (2009), and Getman et al.(2012). The green
stars are Class I objects from Sicilia-Aguilar et al.(2006a) and Reach et al. (2004). The blue X symbols correspond
to X-ray detections consistent with YSO from Getman et al. (2012). The contours mark 10 levels in log scale
from 0.1-2.0 Jy/beam (70\,$\mu$m) and 0.7-10 Jy/beam (160\,$\mu$m). \label{newsource-fig}}
\end{figure*}

Given the strong extended emission throughout the area and the beam dilution for a region located
at 870 pc, only four point sources are clearly identified in the region 
(see Table \ref{source-table} and Figure \ref{70um160um-fig}). Three of them correspond to the known
intermediate-mass Class I candidates 21361836+5728316, 21361942+5728385, and 
21364660+5729384 (Sicilia-Aguilar et al. 2006a),
also known $\delta$, $\epsilon$, and $\alpha$ from Reach et al. (2004). 
The IRAS source 21346+5714 (Sugitani et al. 1991), corresponding to source 
$\gamma$ or 21360798+572637 (Reach et al. 2004; Sicilia-Aguilar et al. 2006a) is marginally
detected at the edge of the image. It appears brighter, redder, and more 
extended than 21364660+5729384, being probably a more massive, less evolved object, but
since it is cut at the edge of the image, we cannot extract any photometric information.
The fourth object is a new source located at 21:37:05.04 +57:30:56.4 (Figure \ref{newsource-fig})
that we name IC1396A-PACS-1. This new source is significantly brighter and more extended 
at 160\,$\mu$m than at 70\,$\mu$m (see Table \ref{source-table}).
The compact part of the source is surrounded by an arc-like extended emission that shows 
a remarkable flux gradient following the contours of the ionized front of the BRC. IC1396A-PACS-1
and its associated arc-like emission clearly dominate the IC\,1396\,A map at 160\,$\mu$m. 
This source was not detected with Spitzer, and the only
previous detection of a source at the same position corresponds to a X-ray source not labeled as 
cluster member due to its low significance (Getman et al. 2012; X-ray source \# 248, with only 3 X-ray net counts). 

\begin{table*}
\caption{Source photometry and integrated luminosities.} 
\label{source-table}
\begin{tabular}{l c c c c c c l}
\hline\hline
Source Name 	    & RA(J2000)   & DEC (J2000) &  F$_{70\mu m}$ (Jy) &  F$_{160\mu m}$ (Jy)  & L$_{integrated}$ (L$_\odot$)  & Class & Other IDs$^*$ \\
\hline
IC1396A-PACS-1      & 21:37:05.04 & +57:30:56.4 &  2.32$\pm$0.23     &  25.5$\pm$2.6        &  1.5$\pm$0.1		   & 0  & \\
21364660+5729384    & 21:36:46.60 & +57:29:38.4 &  3.31$\pm$0.33     &  3.2$\pm$0.3         &  1.00$\pm$0.08		   & I  & $\alpha$ \\
21361942+5728385    & 21:36:19.42 & +57:28:38.5 &  0.97$\pm$0.10     &  5.4$\pm$1.1$^a$      &  0.22$\pm$0.10$^b$  	   & I  & $\epsilon$ \\ 
21361836+5728316    & 21:36:18.36 & +57:28:31.6 &  1.49$\pm$0.15     &  5.4$\pm$1.1$^a$      &  0.26$\pm$0.10$^b$   	   & I  & $\delta$ \\ 
\hline
\end{tabular}
\tablefoot{Photometry of the known cluster members detected by Herschel/PACS in IC\,1396\,A, and
their integrated luminosities (taking into account the ancillary data for the Class I sources, see 
Table \ref{ancillary-table}). The 160\,$\mu$m flux for 21361836+5728316 and 21361942+5728385 corresponds to both
sources, which appear merged in the image, with the peak flux being close to the position of 21361836+5728316.
$^a$ The flux includes both sources, that are merged at 160\,$\mu$m.
$^b$ Luminosities for these sources are very uncertain due to the lack of observations at longer and shorter wavelengths, 
and the fact that the 160\,$\mu$m data point includes both objects.  $^*$ Alternative IDs from Reach et al.(2004).}
\end{table*}

To quantify the fluxes of the sources, we performed aperture photometry with 
IRAF\footnote{IRAF is distributed by the National Optical Astronomy Observatories,
which are operated by the Association of Universities for Research
in Astronomy, Inc., under cooperative agreement with the National
Science Foundation.}.
We selected 6" and 9" apertures for the 70 and 160\,$\mu$m images
and their corresponding aperture corrections of 1.597 and 1.745, according to the PACS manual. 
To avoid potential uncertainties induced by resampling in the final map, the photometry was 
done in the individual fields (large map and mini-map). Due to
the spatially variable cloud emission in IC\,1396\,A, we estimated the sky emission for the photometry by 
measuring in different locations around the sources. The errors were derived considering 
the average sky rms and the correlated noise due to oversampling estimated as 
(3.2/$pixsize$)$^{0.68}$/0.95 and (6.4/$pixsize$)$^{0.73}$/0.88 for
70 and 160~$\mu$m, respectively (Mora private communication). Here, $pixsize$ is the selected
pixel size in the projected
maps (2" for the 70~$\mu$m map and 3" for the 160~$\mu$m map), although taking into account further calibration 
errors and the uncertainties due to strong, highly variable nebular emission result in nominal errors 
of 10\%. 
All Class I sources appear point-like at both wavelengths, but IC1396A-PACS-1 is 
clearly extended at 160\,$\mu$m and may be marginally resolved at 70\,$\mu$m
(see Section \ref{sources}). Therefore, its 160\,$\mu$m flux is likely a lower limit, although
treating it as point-like at 70\,$\mu$m should be accurate.
The photometry of the individual sources is listed in Table \ref{source-table}.

We estimated the total luminosity of the objects by integrating their SEDs. 
Except for the new source IC1396A-PACS-1, the other three objects
had been observed in the near- and mid-IR by 2MASS and/or Spitzer (Table \ref{ancillary-table}), so we 
simply integrated their SEDs including literature data
and the Herschel/PACS data, and extrapolating to modified black body emission
out of the observed wavelength ranges. For 21361836+5728316 and 21361942+5728385, the lack of near-IR
detections and the uncertain 160\,$\mu$m flux (due to source merging and cloud
emission) result in a high uncertainty, although both sources are clearly less luminous
than 21364660+5729384. For 21364660+5729384 we have a more complete dataset and are able to detect 
the turn-down of the SED at high and low frequency, so the estimate is more accurate.

\begin{table}
\caption{Ancillary data for the Class I sources.} 
\label{ancillary-table}
\begin{footnotesize}
\begin{tabular}{l c c c }
\hline\hline
		& 21361836 & 21361942    & 21364660    \\
		& +5728316 & +5728385    & +5729384    \\
\hline
Filter          & Flux (Jy)        & Flux (Jy)           & Flux (Jy) \\
\hline
J		& --- & --- &  $<$4E-4 \\
H		& --- & --- &  9E-4$\pm$1E-4 \\
K		& --- & --- &  5.4E-3$\pm$1E-4 \\
3.6\,$\mu$m 	& 3.2E-3$\pm$4E-4   & 2.9E-3$\pm$3E-4   &  4.3E-2$\pm$4E-3 \\
4.5\,$\mu$m 	& 1.03E-2$\pm$6E-4  & 8.2E-3$\pm$6E-4   &  7.6E-2$\pm$8E-3 \\
5.8\,$\mu$m 	& 2.6E-2$\pm$1E-3   & 1.76E-2$\pm$9E-4  &  0.21$\pm$0.02 \\
8.0\,$\mu$m 	& 3.3E-2$\pm$2E-3   & 2.6E-2$\pm$1E-3   &  0.20$\pm$0.02 \\
24\,$\mu$m 	& 0.346$\pm$4E-3    & 0.355$\pm$0.004   &  1.56$\pm$0.16 \\
\hline
\end{tabular}
\tablefoot{2MASS and Spitzer IRAC/MIPS data for the Class I sources detected with Herschel/PACS. 2MASS data from
Cutri et al. (2003). Spitzer data from Sicilia-Aguilar et al. (2006a).}
\end{footnotesize}
\end{table}

\begin{figure}
\centering
\begin{tabular}{cc}
\epsfig{file=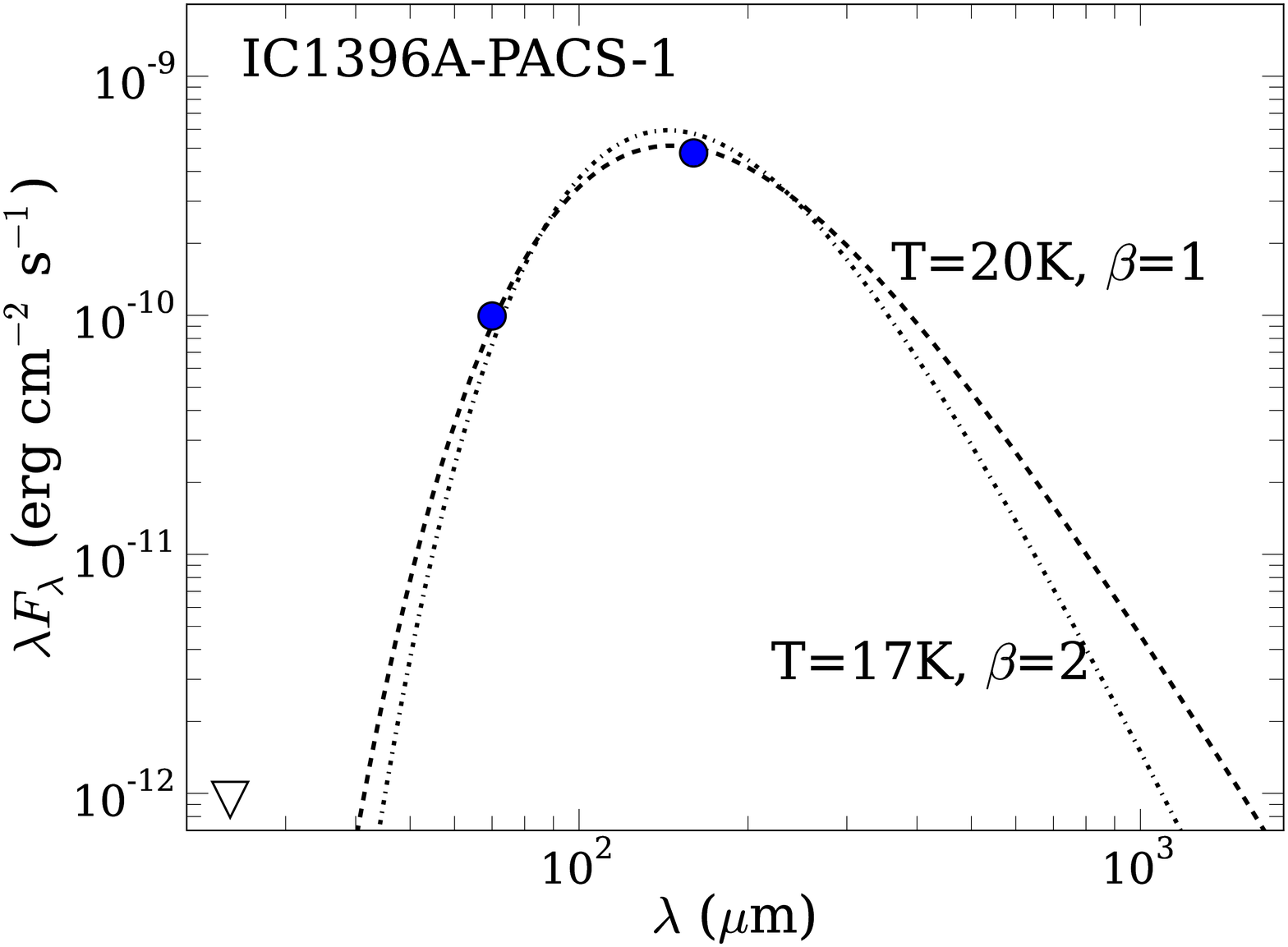,width=0.47\linewidth,clip=} &
\epsfig{file=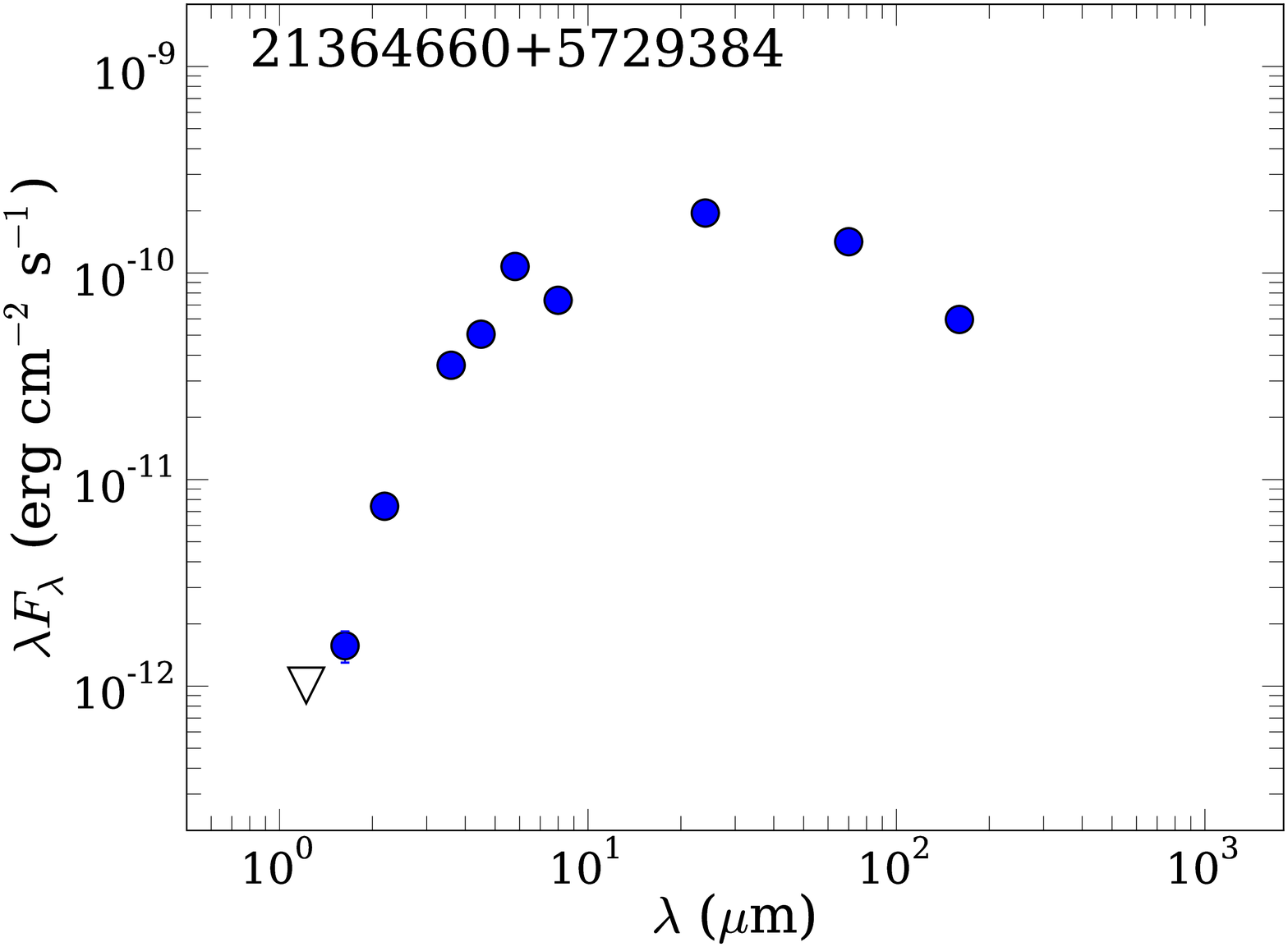,width=0.47\linewidth,clip=}  \\
\epsfig{file=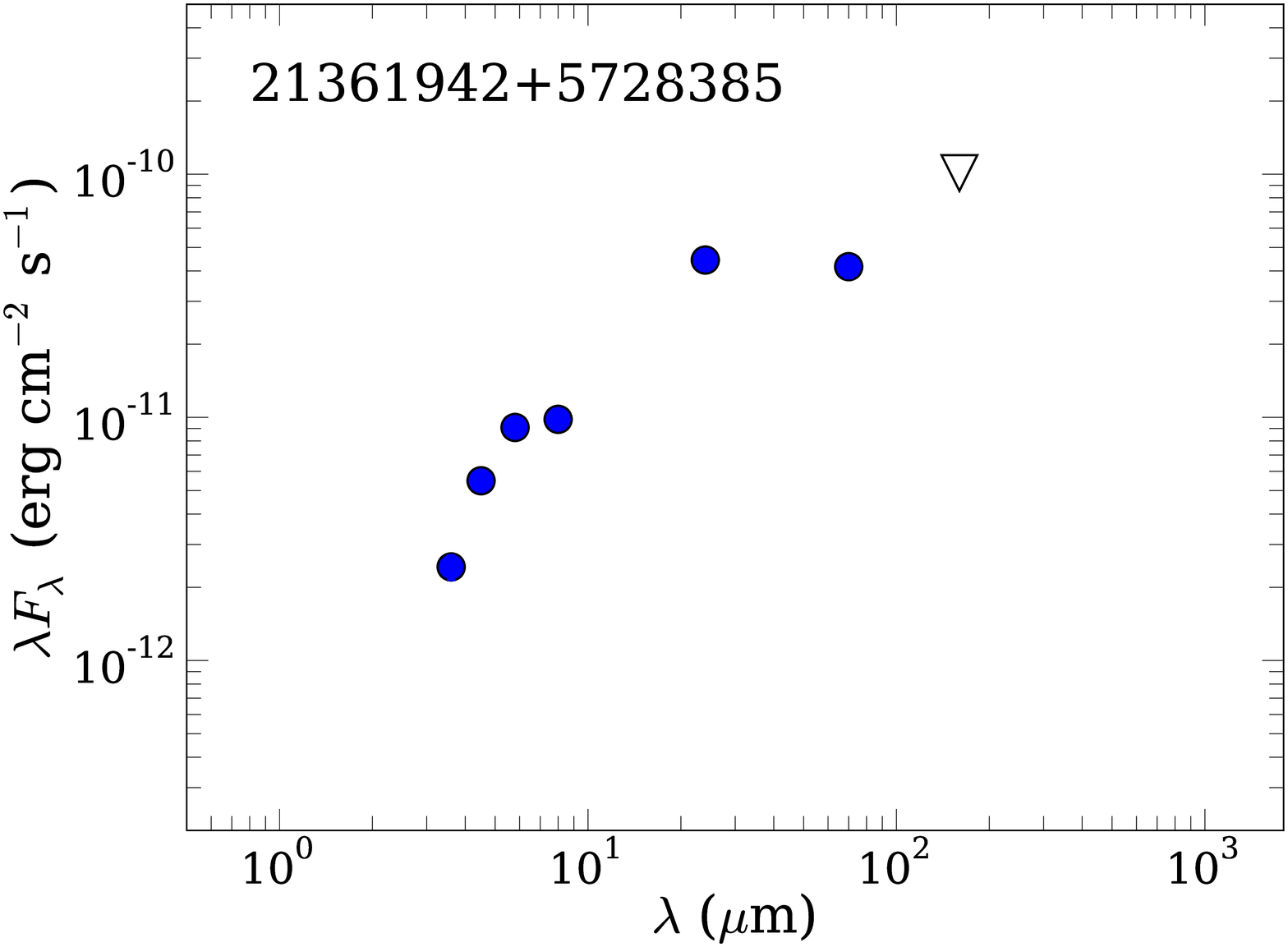,width=0.47\linewidth,clip=}  &
\epsfig{file=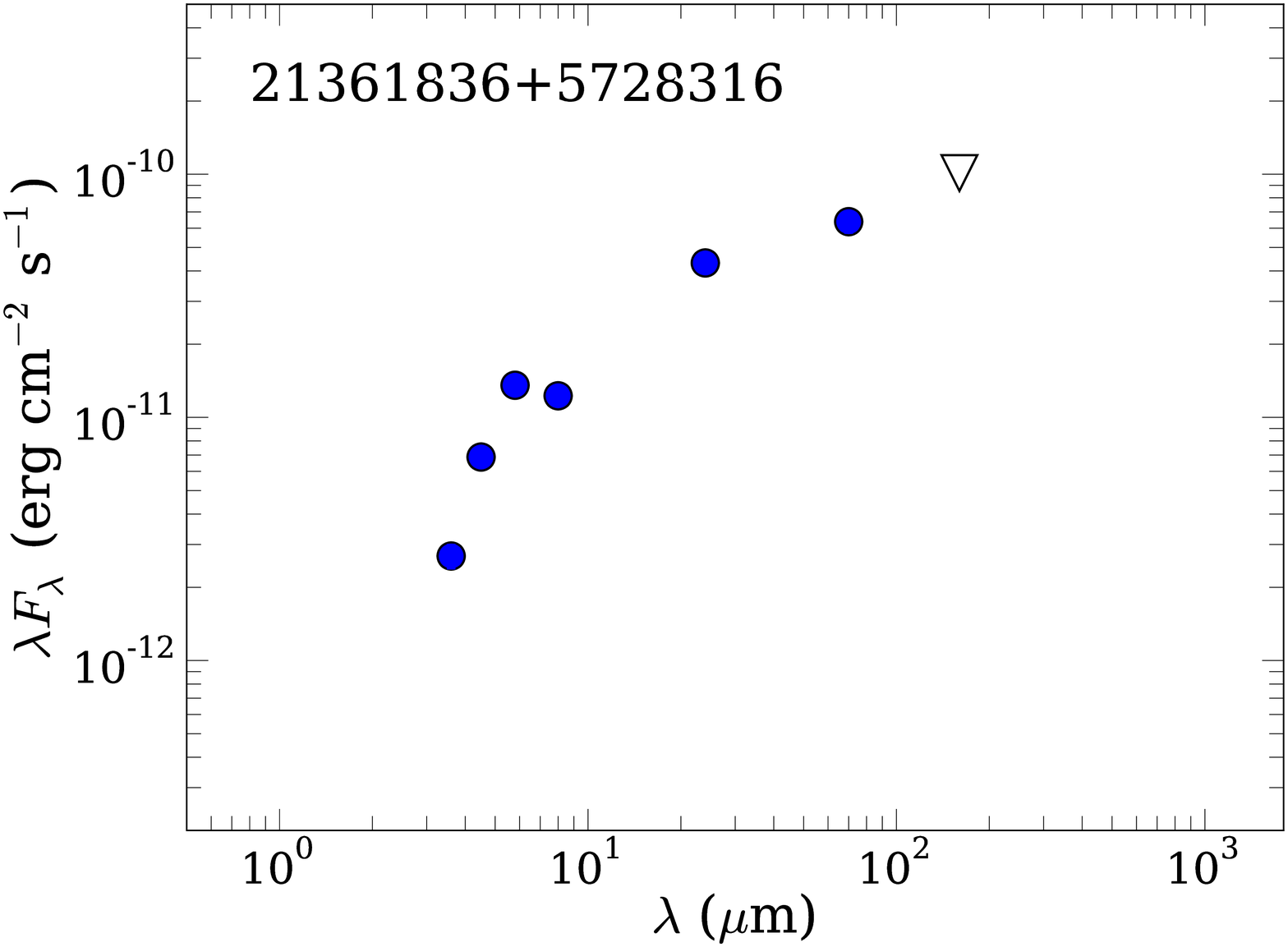,width=0.47\linewidth,clip=} \\
\end{tabular}
\caption{SEDs of the four sources detected in the IC\,1396\,A globule. For IC1396A-PACS-1 we also display
comparative modified black body models with different temperatures and frequency power laws (see text).
Filled circles correspond to detections (errorbars are generally smaller than the points), open inverted
triangles are upper limits.\label{seds-fig}}
\end{figure}

For IC1396A-PACS-1, we follow the procedure
in Sicilia-Aguilar et al. (2013a), based on Ward-Thompson et al.(2002). We assume that the emission
of the Class 0 source can be reproduced by a modified 
black body, given by 
\begin{equation}
	F_\nu = B_{\nu}(T) (1-e^{-\tau_\nu})\Omega. \label{eq1}
\end{equation}
Here, $F_\nu$ is the flux density, $B_{\nu}$(T) is the black body emission for
a temperature T, $\tau_\nu$ is the frequency-dependent optical depth, and $\Omega$ is
the solid angle subtended by the source.
Considering that at long wavelengths,
the optical depth follows a power law with frequency, $\tau_\nu \propto \nu^\beta$,
it is possible to estimate the source temperature,
although for IC1396A-PACS-1 the uncertainty is large due to the fact that we only
have 2 photometric points. Taking the values
from Ward-Thompson et al.(2002) for $\beta$=2 and $\tau_{200\mu m}$=0.06, we obtain a good
fit for a temperature of 17 K, consistent with a pre-stellar core or Class 0 object
(see Figure \ref{seds-fig}). A flatter frequency dependence $\beta$=1 (that could be
related to grain growth) produces a slightly better fit for a temperature of 20 K,
still consistent with the hypothesis of a Class 0 object.  The solid angle
subtended by the object ($\Omega$) is also fitted in this simple model. Models with 
lower T predict larger object radii. In this case, the model with T=17 K, $\beta$=2 would
result in an object larger than the compact source at 70\,$\mu$m ($\sim$9.7"), while the
model with T=20 K, $\beta$=1 is more consistent with a compact source (radius $\sim$6.8"),
even though both models are too simple to offer strong constraints on the object size.

\subsection{Source classification \label{sources}}

Source 21364660+5729384 is brighter at 70\,$\mu$m than at 160\,$\mu$m. Sources 21361836+5728316 and 21361942+5728385 
are merged and surrounded by bright structures at 160\,$\mu$m that make their fluxes very uncertain. Both sources 
are likely brighter at 160\,$\mu$m,  especially  21361942+5728385 (the 160\,$\mu$m 
extended detection peaks at the position of this source). They are thus probably in a more embedded/earlier 
evolutionary phase than 21364660+5729384. Despite the uncertainties caused by merging, 
sources 21361836+5728316 and 21361942+5728385 are
clearly less luminous than 21364660+5729384 at the observed wavelengths.
We traced the SEDs of all objects (Figure \ref{seds-fig}) including all previous
data available in the literature (Table \ref{ancillary-table}), which are consistent with Class I sources,
as previously suggested by Spitzer (Reach et al. 2004; Sicilia-Aguilar et al. 2006). 
The SEDs of these three sources show the characteristic low 8\,$\mu$m emission compared to
the rest of bands, suggestive of a silicate absorption feature, typical of Class I sources.

The SED and far-IR brightness of IC1396A-PACS-1 and the fact that it does not present
any emission at 24\,$\mu$m suggest that it corresponds to a Class 0 or transition Class 0/I
object, the only one known within IC\,1396\,A. 
Given the complexity of the cloud emission, we computed the MIPS 24\,$\mu$m flux 
upper limit based on both the faintest sources detectable on
similar backgrounds, and the sky rms in the position of the object, obtaining for both cases a very
similar result of 3$\sigma$ upper limit 7.5\,mJy at 24\,$\mu$m (see Figure \ref{seds-fig}). This sets a very strong constraint to
the total flux emission of the object at shorter wavelengths. Its relatively high luminosity 
compared to the rest of embedded cloud members suggest it could be the progenitor of an 
intermediate-mass star, although with our resolution we cannot exclude the presence of a small group of objects. 

IC1396A-PACS-1 is clearly extended at 160\,$\mu$m, which would be compatible with it
being a starless core or a Class 0 object, but
the 70\,$\mu$m image reveals a compact source (Figure \ref{newsource-fig}). 
Determining whether the compact source is point-like is more complicated. Its size is
comparable to that of the PACS PSF (FWHM=5.6", beam diameter 6.7"), but the 3-lobe structure of
the PACS PSF, which is clearly visible for the three Class I sources and other point-like
objects in the field, is not clear for IC1396A-PACS-1. This could indicate that the object
is marginally resolved by PACS, although it could be also the result of the strong background
emission around the source. Taking this into account, for a distance of 870 pc, we estimate that
the compact source has a size of $\le$6000 AU.
At 160\,$\mu$m, the source is extended, with an elliptical core slightly more elongated
than the PSF ($>$11.4", $\sim$10$^4$ AU at 870 pc) that merges without clear boundaries into the extended arc-like
rim. At 160\,$\mu$m we would thus observe mostly the envelope emission,
with a typical size close to $\sim$0.1\,pc, but strongly elongated following the cloud contours that are shaped
by the ionization front coming from HD\,206267.
The estimated size at 160\,$\mu$m is consistent with the expectations for Class 0 objects 
(e.g. Lippok et al. 2013), even though it is hard to define the source boundaries at this
wavelength and the whole structure is strongly asymmetric. Further dynamical information
would be needed to determine the extent of the envelope that will collapse onto the central
source(s), as part of the structure may be simply related to the surrounding cloud.

From the compact size and the lack of significant energy output at wavelengths $\le$24\,$\mu$m, 
we conclude that the source is most likely a Class 0 object. 
On the other hand, the marginal ($<$3$\sigma$) X-ray emission found by Getman et al. (2012) 
precisely at the location of the new source is intriguing. If the detected weak X-ray emission were associated
to the source itself, it could point to a later evolutionary stage than Class 0 (e.g., in 
transition between the Class 0 and Class I stages), as
X-ray emission from Class 0 objects is rare and/or very weak (Preibisch 2004; Prisinzano et al. 2008;
Giardino et al. 2006), even though the emission would be still much weaker than observed towards 
typical IC\,1396\,A Class I sources. But if the X-ray emission is associated to
jets in the region (Pravdo et al. 2001; Favata et al. 2002; Sicilia-Aguilar et al. 2008),
we cannot use it as a constrain on the source class.

\subsection{Temperature and cloud structure from Herschel observations \label{temperature}}

Following a similar scheme to that in Sicilia-Aguilar et al. (2013a), we studied the temperature structure of
the cloud. We resampled both images at 70\,$\mu$m and 160\,$\mu$m to the same 3"/pixel scale,
created a new mosaic with the {\sc HIPE} mosaicking tool (resulting in a resampling of 1"/pixel scale),
and derived a ratioed image by dividing the blue by the red one. The temperature
per pixel is then derived based on a modified black body scheme (Preibisch et al. 2013; Roccatagliata et al. 2013). 
This method is well suited to trace the warmer cloud surface temperature. On the other hand, 
SED fitting (using PACS and SPIRE data; Roccatagliata et al. 2013)
computes the beam-averaged dust temperature along the line-of-sight and is more sensitive to the densest
(and thus coolest) central parts of clouds. Roccatagliata et al. (2013) found that the temperatures 
derived from ratioed maps are typically $\sim$5\% higher than the temperatures derived from the
SED fitting. This result is expected because the color temperature
is biased to the warmer cloud surface. In addition, there is a known degeneracy between the dust
color temperature and spectral index (Juvela et al. 2013), which produces different biases
depending on the method used to derive T and $\beta$ from submillimeter observations. In our
case, since we only have observations at 70 and 160\,$\mu$m, our result is approximate.

To construct our temperature maps, we consider that
the emission from each pixel is due to a modified black body (similar to Eq. \ref{eq1}) for a single temperature.
Considering the same solid angle $\Omega$ at both wavelengths,
the ratio of fluxes at 70 and 160\,$\mu$m can be thus written:

\begin{equation}
	\frac{F_{\nu,70}}{F_{\nu,160}} = \frac{B_{\nu,70}(T) (1-e^{-\tau_{\nu,70}})}{B_{\nu,160}(T) (1-e^{-\tau_{\nu,160}})}. \label{eq2}
\end{equation}

\begin{figure*}
\centering
\epsfig{file=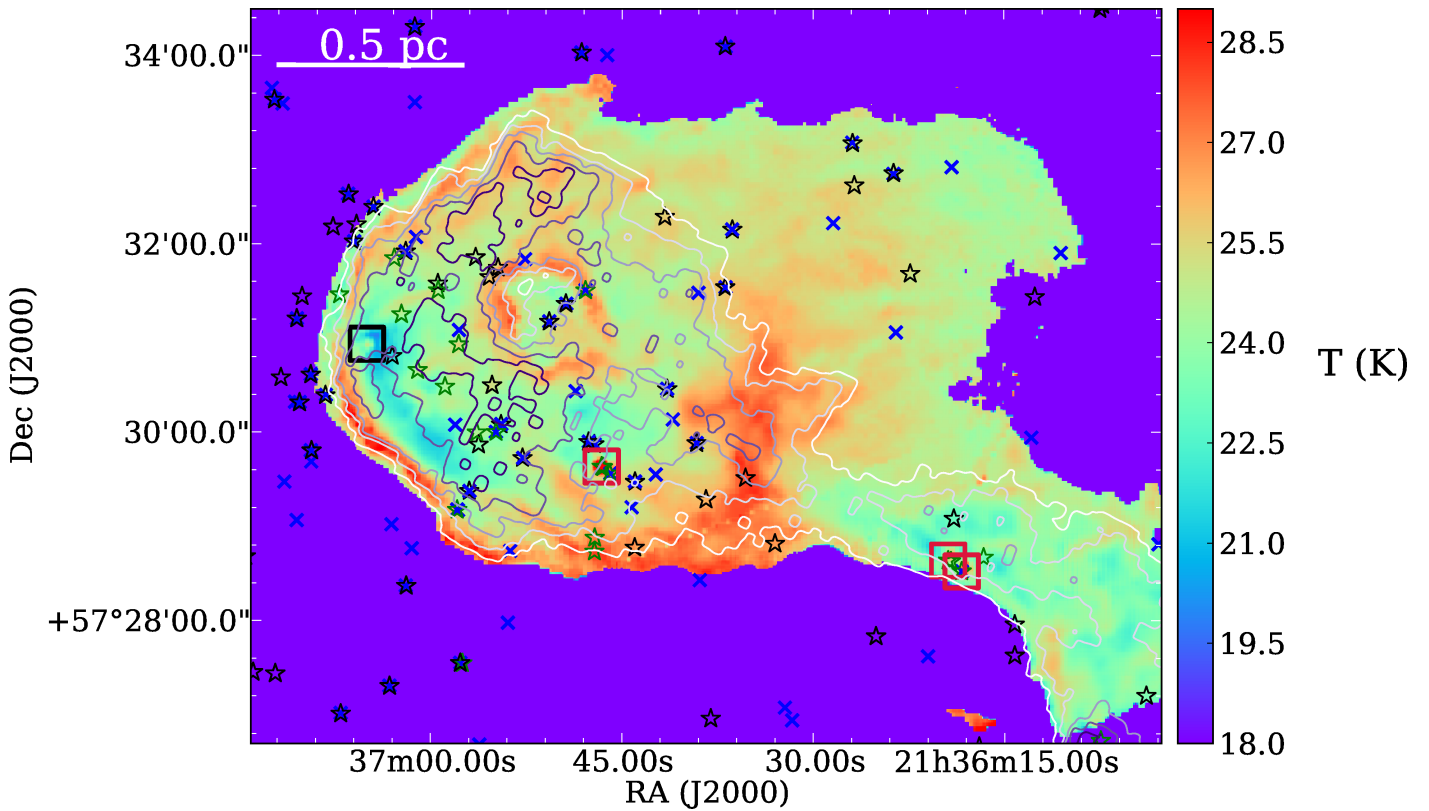,width=1.0\linewidth,clip=}
\caption{Temperature map for IC\,1396\,A (see text). The contours mark the 10, 20, 30, 40, 50 K km s$^{-1}$
levels of the integrated CO(3-2) emission from archival JCMT data. The black stars denote Class II and Class III objects
from Sicilia-Aguilar et al. (2005, 2006a,b), Morales-Calder\'{o}n et al. (2009), and Getman et al.(2012). The green
stars are Class I objects from Sicilia-Aguilar et al.(2006a) and Reach et al. (2004). The blue X symbols correspond
to X-ray detections consistent with YSO from Getman et al. (2012). The objects discussed in this work are marked as
large squares (black for Class 0 objects, red for Class I objects). Note that due to the exclusion of pixels with low
fluxes (see text), the only significant temperatures are those inside the globule and not those in the external
parts that appear violet in the image. \label{temperature-fig}}
\end{figure*}

If we take $\beta$=2,
the flux ratio between two wavelengths is thus a function of T alone, which can be determined, tracing the
temperature structure from the cloud (Figure \ref{temperature-fig}). For this exercise, we excluded the pixels
with fluxes below 0.0014 Jy/pix (at 70\,$\mu$m) and below 0.0015 Jy/pix (at 160\,$\mu$m; for
the {\sc HIPE}-resampled pixel size 1") to avoid uncertainties due to the typical image noise. Therefore, 
the only significant values of the temperature are those inside the globule.
The surroundings of the IC\,1396\,A, corresponding to the H II region around the cluster
Tr\,37, have uncertain fluxes due to the nearby bright globule and to
the low background, which can lead to some unphysical temperature distributions out of the BRC.

A comparison of the temperature structure with the location of the known Class I/II/III 
sources reveals that the sources are on slightly warmer (T$>$21 K), net-like structures throughout 
the cloud. Part of this may be the effect of beam dilution on faint, distant sources with far-IR 
emission. Since not all warm structures observed in the cloud have associated Spitzer/X-ray sources, 
there is also the possibility that embedded sources are also
heating their surroundings, in a similar way to what we observe around V~390~Cep. The main
difference would be the lower mass (and energetic output) of most of the Class I/II/III 
sources in the globule. Warmer, sourceless areas may be associated with external heating 
by HD\,206267.  Cloud heating by low-mass sources is also seen in low-mass regions like 
the Coronet cluster (Sicilia-Aguilar et al. 2013a).

IC1396A-PACS-1 is surrounded by a cold area, but its central part is
clearly hotter, revealing incipient star formation. A comparison of the temperature
map with archival JCMT CO(3-2) data\footnote{http://www4.cadc-ccda.hia-iha.nrc-cnrc.gc.ca/data/pub/JCMT}
shows that the CO emission is correlated with the known Class I/II/III sources and the warmest
areas of the globule, while the area around IC1396A-PACS-1 does not have any significant CO emission
above the plateau level observed in the coldest areas of the cloud. 
The archival data corresponds to CO(3-2) fluxes integrated over a 0.9 GHz window, so it
not possible to obtain any significant dynamical information of the region nor its objects.
The lack of CO emission is consistent with the low temperatures in the area around IC1396A-PACS-1.
Even though CO emission from outflows would be expected for a Class 0/I object, the lack of velocity
resolution of the available JCMT data together with the strong cloud emission do not allow us to distinguish
potential source-related emission. Starless cores are typically affected by  
CO depletion (Caselli et al. 1999; Bergin et al. 2002; Nielbock et al. 2012; 
Lippok et al. 2013), but IC1396A-PACS-1 is too compact and too bright to be considered
starless. Further observations, especially velocity-resolved, would be needed to confirm whether
the object presents distinct CO emission. In case of a CO depleted object, taking into 
account the critical number density values for CO depletion 
(n$>$10$^5$~cm$^{-3}$; Caselli et al. 1999) and a minimum size of the core as seen at 160\,$\mu$m 
($\sim$15"$\times$20", or $\sim$13000~AU$\times$17400~AU as seen in projection), we would expect 
a mass of nearly 10~M$_\odot$ in the core. Considering the uncertainties in the object size and that 
a distinct, compact object is resolved at 70\,$\mu$m (so the density is likely higher in the center), this would be in
agreement with the object being the progenitor of an intermediate-mass star or a small stellar group
(as observed, for instance, in highly fragmented clouds like the Coronet cluster; Sicilia-Aguilar et al. 2013a).

\section{Discussion: Sequential and triggered star formation in Tr\,37\label{discussion}}

The cometary structure of the BRC IC\,1396\,A has since long been identified as a 
region shaped by the star HD\,206267 (Loren et al. 1975) and a potential site for 
triggered star formation (Sugitani et al. 1991; Patel et al. 1995; Reach et al. 2004; 
Sicilia-Aguilar et al. 2005, 2006a; Morgan et al. 2009, 2010; Getman et al. 2012). 
The new Class 0 object IC1396A-PACS-1 at the tip of IC\,1396\,A
is a sign that the star formation process is not yet finished in the region. 
We do not find evidence 
of any further star formation, even though due to contrast problems with the cloud 
emission, we cannot exclude the presence of further low-luminosity
Class 0 objects. Nevertheless, the temperature map  
(Figure \ref{temperature-fig}) suggests that only the tip of the BRC behind the
ionization rim has temperatures low enough to allow the formation of new 
stars, in agreement with archival CO JCMT observations. 

IC\,1396\,A appears consistent with what one would expect from RDI triggered 
star formation. The warmer outer dust layer can be a tracer of the ionization front 
(similar to Figure 2 in Bisbas et al. 2011), and the cold, extended structure that 
harbors IC1396A-PACS-1 lies right behind the ionization front. This would be in 
agreement with the general/qualitative RDI picture where dense cores form at the 
edges of the cloud between the ionization front and the compression shock. However, 
there is a substantial difference between the observations of IC\,1396\,A and the 
simulations of Bisbas et al. (2011) and others: they model small 5-10~M$_\odot$ 
cloudlets. These small cloudlets are expected to form a few low-mass stars within $<$1~Myr 
years before being completely ablated and photoevaporated. In contrast, IC\,1396\,A is 
a much larger globule, comparable to others like IC~1396~N and Cep~B ($>$200\,M$_\odot$;  
Getman et al. 2007, 2009, 2012). In these larger regions, the triggered star formation 
does not appear restricted to a single short episode, 
but is rather a multi-episodic (or continuous) process that can last for many Myr years.

With the typical velocity dispersal observed in the low-mass members of Tr\,37 
(Sicilia-Aguilar et al. 2006b), stars are expected to move from their formation 
sites at a rate of about 1-2 pc per Myr. This is comparable to the size of the 
IC\,1396\,A globule for the age of Class I systems found within it. Nevertheless, 
the lack of Class I sources out of IC\,1396\,A
and the presence of small, distinct mini-clusters within Tr\,37 suggest 
that for ages about 1 Myr, dynamical evolution is not yet important in this region.
These mini-clusters consist of compact ($<$0.5\,pc) groups of a few (2-10) stars with higher 
IR excesses and accretion rates than the main cluster. Sometimes they are also associated to small patches of nebulosity, 
all suggesting that they are younger than the surrounding, more evolved population 
(Barentsen et al. 2011; Getman et al. 2012; Sicilia-Aguilar et al. 2013b).
These mini-clusters could be a sign that star formation does not progress in a
uniform way and several episodes of star formation may occur, even at scales as small as 0.5\,pc,
suggesting "clumpy" or multi-episodic star formation within the same cloud.
The formation of IC1396A-PACS-1 at the tip of the cloud and surrounded by younger 
Class I/II objects may also explain the origin of the mini-clusters. 
Getman et al. (2009) proposed two distinct scenarios for RDI star formation: slow shock 
propagation through the globule (where stellar kinematics and dynamical drift is important), 
and fast shock propagation (where stellar kinematics plays a less important role). The
star formation efficiency in the Cep~B cloud supported the second 
scenario, which seems also to be the case for IC\,1396\,A.

IC1396A-PACS-1 fits in this picture of continuous/multi-episodic
triggered star formation. Given the age and evolutionary state differences between IC1396A-PACS-1 and the rest of the
IC\,1396\,A population, other possibilities could be spontaneous star formation or triggering
related to outflows from the previous population in the cloud. However, the position of the Class 0 source and the
arc-shaped structure with a very steep gradient at far-IR wavelengths, as seen with Herschel/PACS, 
suggests some causal link between the ionization front and the object.  
Our [S\,II] imaging of IC\,1396\,A did not reveal any outflows in the surroundings of IC1396A-PACS-1,
despite finding strong forbidden line emission at the ionization front, at the edge of the bubble created
by V~390~Cep (especially, near the low-mass star 14-141), and some weaker outflow emission probably 
associated with 21364660+5729384 or other 
of the lower-mass Class I objects (Sicilia-Aguilar et al. 2013b).
The Herschel data, together with our previous observations, 
thus favor the scenario where several distinct episodes of sequential or triggered star formation
have occurred. The age and evolutionary stage gradient observed between the Tr\,37 cluster and the Class I/II 
population associated with IC\,1396\,A (Sicilia-Aguilar et al. 2005, 2006a; Barentsen et al. 2011; Getman et al. 2012)
shows that, after formation of the main Tr\,37 cluster approximately 4 Myr ago, a second generation of stars was born
to the west, only $\sim$1-2 Myr ago, and star-building is still active in the cloud. 
The Class I/II objects are located in the less dense parts of the cloud, 
maybe because most of the original material is now dispersed or in the process of being dispersed, after suffering 
irradiation from HD\,206267 and from the new generation of YSO. 

Although most of the Spitzer-identified Class I/II objects within IC\,1396\,A cannot be
individually resolved with PACS due to their lower luminosity and the high cloud background, the Herschel observations
suggest that these low-mass stars may be responsible for heating and dispersal of the cloud at small scales. 
These objects correspond mostly to solar-type and low-mass stars/protostars (below 2\,M$_\odot$ and down to $\sim$0.1\,M~$_\odot$;
Sicilia-Aguilar et al. 2005, 2013b; Getman et al. 2012) and appear spatially related to the warmest parts
of the cloud  (Figure \ref{temperature-fig}). HD\,206267 could be also the heating source for the
warm structure observed around 21:36:36, 57:30:30. In this case, we would expect this structure to be
raised towards the observer, receiving full illumination by the O6.5 star.
Only the denser part of the cloud, where the 
column density is higher and the temperature lower, would have started its collapse very recently, given rise
to the Class 0 object IC1396A-PACS-1. If the cloud heating and dispersal continue as observed, this object may be the last
one formed in the region, and may evolve into a structure similar to $\gamma$/21360798+572637 once the less dense parts of the
cloud have been eroded. 
Nevertheless, considering that temperature maps are
dominated by the higher temperatures in the region, our maps of the IC\,1396\,A region may be
affected by the hotter, ionized, cloud surface, that is directly exposed to the action of the
massive and intermediate stars in Tr\,37. Further observations (e.g. high spatial-resolution
line observations) would be required to exclude the
presence of undetected cold and dense clumps that may continue to form stars in the region.

Including the observed IR and X-ray sources, the stellar density of the cloud is much lower than found
in sparse associations like the Coronet cluster (Sicilia-Aguilar et al. 2013a), even after correcting
for completeness and distance differences. The Coronet cloud is substantially cooler than the IC\,1396\,A region due 
to the lack of external ionizing sources. The heating in the embedded parts of the Coronet appears to
be exclusively related to the intermediate- and low-mass sources within the cloud.
If star formation is halted by external cloud heating and dispersal, the initial conditions of Class I and Class II objects in
and around IC\,1396\,A will be very different from those observed in the sparse associations, which could also
affect the subsequent disk evolution. 

In case of regions with a larger number of OB stars, their effect over the remnant cloudlets may
be more extreme than in IC\,1396\,A. For instance, the Carina Nebula Complex extends over more 80\,pc 
in the sky and it contains at least 65 O-type stars and four Wolf-Rayet stars.
The average temperature of most of the complex is about 30 K, while  at the edge of the nebula 
the clouds temperature decrease down to 26 K (Roccatagliata et al 2013).
The hotter parts of the cloud are located in the central region 
around $\eta$ Car ($\sim$25\,pc$\times$25\,pc), which hosts the young clusters Trumpler 14, 15
and 16 and contains about 80\% of the high-mass stars of the entire complex.
The temperatures are between 30 and 50 K (excluding the position
of $\eta$ Car itself, which reaches values of 60 K).
The local temperature of the cloud is 
related to the number of high-mass stars in the cluster. 
In such environments, the potential for several sequences of star formation would be expected to
be more limited, due to the extreme disruption of cloudlets and the higher temperatures of the
surrounding environment.

\section{Summary \label{summary}}

We present the first results of our Open Time Herschel proposal on the Tr\,37 cluster.
We used PACS observations at 70 and 160\,$\mu$m to trace the BRC IC\,1396\,A,  
previously known to host a population of T Tauri and Class I objects
with ages younger than the main Tr\,37 cluster. Our results can be summarized as:

\begin{itemize}
\item We identify for the first time a Class 0 object within the IC\,1396\,A globule, 
labeled as IC1396A-PACS-1. It is the only Class 0 candidate
known in the region. This object was not detected with Spitzer, although there is a low-significance
X-ray detection that could be related to it. The object
is located in the coldest, CO-depleted region of the cloud, directly behind the ionized rim.
Its estimated luminosity ($\sim$1.5$\pm$0.1 L$_\odot$)
suggest that it is the precursor of an intermediate-mass star or a small stellar group.

\item Three of the Class I sources in IC\,1396\,A are also detected with Herschel/PACS,
completing their Spitzer and 2MASS data, and confirming their classification. Their luminosities
suggest that they are precursors of solar-mass to intermediate-mass stars.

\item The discovery of the new Class 0 object IC1396A-PACS-1 suggest 
multi-episodic star formation in the region. Previous episode(s)
would have given birth to the embedded Class I and Class II objects in IC\,1396\,A,
while a later one would have originated IC1396A-PACS-1 at the densest and coldest
part of the cloud. The position of IC1396A-PACS-1 with respect to the ionization
front observed in the BRC is consistent with triggered star formation via RDI.

\item The presence of two populations with different ages and resulting
from distinct star formation episodes within the IC\,1396\,A cloud may also
explain the mini-clusters observed in Tr\,37. These mini-clusters consist of small
groups of stars, sometimes surrounded by nebulosity, that appear younger than
the main Tr\,37 population. Mini-clusters could result from clumpy or multi-episodic
star formation within the same cloud, as is observed in IC\,1396\,A.

\item The temperature map of IC\,1396\,A shows that heating in the cloud 
responds to two mechanisms: strong external heating by the O6.5 system HD\,206267, and 
to a lesser extent in the most embedded parts, localized heating
by stars and protostars within the cloud. The most massive stars (e.g. V~390~Cep) also
contribute to active dispersal of the cloud material. Heating and dispersal may compromise
future star formation in the warmest regions, and external heating from
massive sources may also result in differences between clusters containing OB stars and
sparse, low-mass associations. Comparing IC\,1396\,A to other regions
suggests that the potential for multi-episodic star formation in a cloud 
may be strongly dependent on the environment and the effect of the first-formed
stars, with OB stars having a dramatic role in heating and dispersal
of the cloud.

\end{itemize}

\vskip 0.5truecm
Acknowledgments: We thank Bruno Altieri from the Herschel Helpdesk for his valuable help with the 
data reduction, and Lorenzo Piazzo for making available the Unimap code. We also thank 
Sofia Sayzhenkova from the computing support at the Departamento de F\'{i}sica
Te\'{o}rica, and the referee and the Editor, M. Walmsley, for their useful comments that helped to improve this paper.
A.S.A. acknowledges support by the Spanish MICINN/MINECO "Ram\'{o}n y Cajal" 
program, grant number RYC-2010-06164,
and the action ``Proyectos de Investigaci\'{o}n fundamental no orientada", grant number
AYA2012-35008. C.E. is partly supported by Spanish MICINN/MINECO
grant AYA2011-26202. V.R. is supported by the DLR grant number 50 OR 1109
and by the {\it Bayerischen Gleichstellungsf\"orderung} (BGF).


\begin{thebibliography}{}

\bibitem[Abt(1986)]{1986ApJ...304..688A} Abt, H.~A.\ 1986, \apj, 304, 688 

\bibitem[Arzoumanian et al.(2011)]{2011A&A...529L...6A} Arzoumanian, D., Andr{\'e}, P., Didelon, P., et al.\ 2011, \aap, 529, L6 

\bibitem[Barentsen et al.(2011)]{2011MNRAS.415..103B} Barentsen, G., Vink, J.~S., Drew, J.~E., et al.\ 2011, \mnras, 415, 103 

\bibitem[Bergin et al.(2002)]{2002ApJ...570L.101B} Bergin, E.~A., Alves, J., Huard, T., \& Lada, C.~J.\ 2002, \apjl, 570, L101 

\bibitem[Bisbas et al.(2011)]{2011ApJ...736..142B} Bisbas, T.~G., W{\"u}nsch, R., Whitworth, A.~P., Hubber, D.~A., \& Walch, S.\ 2011, \apj, 736, 142 

%\bibitem[Bisbas et al.(2012)]{2012MNRAS.427.2100B} Bisbas, T.~G., Bell, T.~A., Viti, S., Yates, J., \& Barlow, M.~J.\ 2012, \mnras, 427, 2100 

\bibitem[Caselli et al.(1999)]{1999ApJ...523L.165C} Caselli, P., Walmsley, C.~M., Tafalla, M., Dore, L., \& Myers, P.~C.\ 1999, \apjl, 523, L165 

\bibitem[Chen \& Huang(2010)]{2010RAA....10..777C} Chen, S., \& Huang, M.\ 2010, Research in Astronomy and Astrophysics, 10, 777 

\bibitem[Contreras et al.(2002)]{contreras02}  Contreras, M.E., Sicilia-Aguilar, A., Muzerolle, J., Calvet, N., Berlind, P., Hartmann, L. 2002, AJ, 124, 1585

\bibitem[Elmegreen(1998)]{1998ASPC..148..150E} Elmegreen, B.~G.\ 1998, Origins, 148, 150 

\bibitem[Favata et al.(2002)]{favata02} Favata, F.; Fridlund, C. V. M.; Micela, G.; Sciortino, S.; Kaas, A. A., 2002, ASPC 277, 467

\bibitem[Getman et al.(2007)]{2007ApJ...654..316G} Getman, K.~V., Feigelson, E.~D., Garmire, G., Broos, P., \& Wang, J.\ 2007, \apj, 654, 316 

\bibitem[Getman et al.(2009)]{2009ApJ...699.1454G} Getman, K.~V., Feigelson, E.~D., Luhman, K.~L., et al.\ 2009, \apj, 699, 1454 

\bibitem[Getman et al.(2012)]{2012MNRAS.426.2917G} Getman, K.~V., Feigelson, E.~D., Sicilia-Aguilar, A., et al.\ 2012, \mnras, 426, 2917 

\bibitem[Giardino et al.(2006)]{2006A&A...453..241G} Giardino, G., Favata, F., Silva, B., et al.\ 2006, \aap, 453, 241 

\bibitem[Juvela et al.(2013)]{2013A&A...556A..63J} Juvela, M., Montillaud, J., Ysard, N., \& Lunttila, T.\ 2013, \aap, 556, A63 

\bibitem[Kessel-Deynet \& Burkert(2003)]{2003MNRAS.338..545K} Kessel-Deynet, O., \& Burkert, A.\ 2003, \mnras, 338, 545 

\bibitem[Lippok et al.(2013)]{2013arXiv1308.2958L} Lippok, N., Launhardt, R., Semenov, D., et al.\ 2013, A\&A 560, A41

\bibitem[Loren et al.(1975)]{1975ApJ...195...75L} Loren, R.~B., Peters, W.~L., \& Vanden Bout, P.~A.\ 1975, \apj, 195, 75 

\bibitem[Marschall \& van Altena(1987)]{marschall87} Marschall, L.A. \& van Altena, W.F., 1987, \aj, 94,71

\bibitem[Mercer et al.(2009)]{2009AJ....138....7M} Mercer, E.~P., Miller, J.~M., Calvet, N., Hartmann, L., Hernandez, J., Sicilia-Aguilar, A., \& Gutermuth, R.\ 2009, \aj, 138, 7 

\bibitem[Miao et al.(2009)]{2009ApJ...692..382M} Miao, J., White, G.~J., Thompson, M.~A., \& Nelson, R.~P.\ 2009, \apj, 692, 382 

\bibitem[Mookerjea et al.(2012)]{2012A&A...542L..17M} Mookerjea, B., Ossenkopf, V., Ricken, O., et al.\ 2012, \aap, 542, L17 

\bibitem[Morales-Calder{\'o}n et al.(2009)]{2009ApJ...702.1507M} Morales-Calder{\'o}n, M., Stauffer, J.~R., Rebull, L., et al.\ 2009, \apj, 702, 1507 

\bibitem[Morgan et al.(2009)]{2009MNRAS.400.1726M} Morgan, L.~K., Urquhart, J.~S., \& Thompson, M.~A.\ 2009, \mnras, 400, 1726 

\bibitem[Morgan et al.(2010)]{2010MNRAS.408..157M} Morgan, L.~K., Figura, C.~C., Urquhart, J.~S., \& Thompson, M.~A.\ 2010, \mnras, 408, 157 

\bibitem[Nielbock et al.(2012)]{2012A&A...547A..11N} Nielbock, M., Launhardt, R., Steinacker, J., et al.\ 2012, \aap, 547, A11 

\bibitem[Ogura et al.(2007)]{2007PASJ...59..199O} Ogura, K., Chauhan, N., Pandey, A.~K., et al.\ 2007, \pasj, 59, 199 

\bibitem[Ott(2010)]{2010ASPC..434..139O} Ott, S.\ 2010, Astronomical Data Analysis Software and Systems XIX, 434, 139 

\bibitem[Patel et al.(1995)]{pat95}  Patel, N.A. , Goldsmith, P.F., Snell, R.L., Hezel, T. \& Xie, T., 1995, ApJ , 447, 721

\bibitem[Patel et al.(1998)]{pat98} Patel, N.A.,  Goldsmith, P.F., Heyer, M.H. \& Snell, R.L., 1998, ApJ , 507, 241

\bibitem[Peter et al.(2012)]{2012A&A...538A..74P} Peter, D., Feldt, M., Henning, T., \& Hormuth, F.\ 2012, \aap, 538, A74 

\bibitem[Piazzo et al.(2012)]{unimap}  L. Piazzo, D. Ikhenaode, P. Natoli, M. Pestalozzi , F. Piacentini and A. Traficante: "Artifact removal for GLS map makers by means of post-processing", IEEE Trans. on Image Processing, Vol. 21, Issue 8, pp. 3687-3696, 2012. 

\bibitem[Piazzo et al.(2013)]{unimap2}  L. Piazzo, 2013,  http://arxiv.org/abs/1301.1246

\bibitem[Pilbratt et al.(2010)]{2010A&A...518L...1P} Pilbratt, G.~L., Riedinger, J.~R., Passvogel, T., et al.\ 2010, \aap, 518, L1 

\bibitem[Platais et al.(1998)]{pla98} Platais, I,  Kozhurina-Platais, V., van Leeuwen, F., 1998, \aj, 116, 2423

\bibitem[Poglitsch et al.(2010)]{2010A&A...518L...2P} Poglitsch, A., Waelkens, C., Geis, N., et al.\ 2010, \aap, 518, L2 

\bibitem[Pravdo et al.(2001)]{pravdo01} Pravdo, S. H.; Feigelson, E. D.; Garmire, G.; Maeda, Y.; Tsuboi, Y.; Bally, J., 2001, Nature 413, 708

\bibitem[Preibisch(2004)]{2004A&A...428..569P} Preibisch, T.\ 2004, \aap, 428, 569 

\bibitem[Preibisch et al.(2013)]{preibisch13} Preibisch, T. and Roccatagliata, V. and Gaczkowski, B. and Ratzka, T, 2012, \aap, A132, 541

\bibitem[Prisinzano et al.(2008)]{2008ApJ...677..401P} Prisinzano, L., Micela, G., Flaccomio, E., et al.\ 2008, \apj, 677, 401 

\bibitem[Reach et al.(2004)]{rea04} Reach, W., Rho, J., Young, E., et al. 2004, \apjs, 154, 385

\bibitem[Roccatagliata et al.(2013)]{roccatagliata13} Roccatagliata, V. and Preibisch, T. and Ratzka, T. and Gaczkowski, B.,  2013, \aap A6, 554

\bibitem[Sicilia-Aguilar et al. (2004)]{sicilia04}Sicilia-Aguilar, A., Hartmann, L., Brice\~{n}o, C., Muzerolle, J.,Calvet, N., 2004, AJ 128, 805

\bibitem[Sicilia-Aguilar et al. (2005)]{sicilia05} Sicilia-Aguilar, A., Hartmann, L., Hern\'{a}ndez, J., Brice\~{n}o, C., Calvet, N., 2005, AJ 130, 188

\bibitem[Sicilia-Aguilar et al.(2006a)]{sicilia06ir} Sicilia-Aguilar, A., Hartmann, L., Calvet, N., Megeath, S.T., Muzerolle, J., Allen, L., D'Alessio, P., Mer\'{\i}n, B., Stauffer, J., Young, E., Lada, C., 2006a,ApJ  638, 897

\bibitem[Sicilia-Aguilar et al.(2006b)]{sicilia06opt} Sicilia-Aguilar, A., Hartmann, L., F\"{u}r\'{e}sz, G., Henning, Th., Dullemond, C., Brandner, W., 2006b, AJ 132, 2135

\bibitem[Sicilia-Aguilar et al.(2008)]{sic08cra}  Sicilia-Aguilar, A.; Henning, Th.; Juh\'{a}sz, A.; Bouwman, J.; Garmire, G.; Garmire, A., 2008, \apj, 687, 1145

\bibitem[Sicilia-Aguilar et al.(2011b)]{2011ApJ...742...39S} Sicilia-Aguilar, A., Henning, T., Dullemond, C.~P., et al.\ 2011b, \apj, 742, 39 

\bibitem[Sicilia-Aguilar et al.(2013a)]{2013A&A...551A..34S} Sicilia-Aguilar, A., Henning, T., Linz, H., et al.\ 2013a, \aap, 551, A34 

\bibitem[Sicilia-Aguilar et al.(2013b)]{lowmass}  Sicilia-Aguilar, A., Kim, J.S., Sobolev, A., Getman, K., Henning, Th., Fang, M., 2013b, A\&A 559, 29

\bibitem[Siess et al.(2000)]{sies00} Siess, L., Dufour, E. \&  Forestini, M. 2000 A\&A , 358, 593

\bibitem[Simonson \& van Someren Greve(1976)]{sim76} Simonson, S.C. \&  van Someren Greve, H.V.,1976, A\&A, 46, 261

\bibitem[Sugitani et al.(1991)]{1991ApJS...77...59S} Sugitani, K., Fukui, Y., \& Ogura, K.\ 1991, \apjs, 77, 59 

\bibitem[Sugitani \& Ogura(1994)]{1994ApJS...92..163S} Sugitani, K., \& Ogura, K.\ 1994, \apjs, 92, 163 

\bibitem[Ward-Thompson et al.(2002)]{2002MNRAS.329..257W} Ward-Thompson, D., Andr{\'e}, P., \& Kirk, J.~M.\ 2002, \mnras, 329, 257 

\bibitem[Weikard et al.(1996)]{1996A&A...309..581W} Weikard, H., Wouterloot, J.~G.~A., Castets, A., Winnewisser, G., \& Sugitani, K.\ 1996, \aap, 309, 581 

\end{thebibliography}
\end{document}